\documentclass[%
 reprint,
 amsmath,amssymb,
 aps,
pra,
]{revtex4-1}

\usepackage{graphicx}
\usepackage{dcolumn}
\usepackage{bm}

\usepackage{amsfonts}
\usepackage{amssymb}

\begin{document}




\title{Conditional Spectroscopy via Non-Stationary Optical Homodyne Quantum State Tomography}



\author{Johannes Thewes}
\author{Carolin L\"uders}%
\author{Marc A{\ss}mann}
 \email{marc.assmann@tu-dortmund.de}
\affiliation{%
 Experimentelle Physik 2,
	Technische Universit\"at Dortmund,
	D-44221 Dortmund, Germany
}%

\date{\today}







\begin{abstract}
Continuous variable quantum state tomography is one of the most powerful techniques to study the properties of light fields in quantum optics. However, the need for a fixed phase reference has so far prevented widespread usage in other fields such as semiconductor spectroscopy. Here, we introduce non-stationary quantum state tomography, which adapts the technique to the special requirements of ultrafast spectroscopy. In detail, we gain access to the amplitude and phase of light fields with a temporal resolution of about 100\,fs without the need for a fixed phase reference. Further, we show how our technique allows us to perform conditional studies of stochastic dynamics that are inaccessible experimentally by conventional means, and demonstrate the capabilities experimentally by monitoring the stochastic dynamics of a thermal light field on the sub-ps scale. Finally, we discuss differences and similarities to more standard Hanbury Brown-Twiss photon correlation experiments, which may be considered as the discrete variable analogues of our technique.
\end{abstract}

\maketitle


\section{Introduction}

Optical spectroscopy, e.g. in semiconductor physics, follows a special approach towards characterizing light fields. Here, the light field is used as a messenger to gather as much information as possible about some system that has interacted with the light field via light-matter interaction. Accordingly, it is not the state of the light field itself that is of fundamental interest, but instead the matter system, e.g. one investigates a semiconductor sample via its photoluminescence. Preferably, one aims at resolving the system dynamics with high temporal resolution on the sub-ps level. This poses two major challenges with respect to the detection schemes used. First, commonly used detectors in semiconductor spectroscopy, such as photo diodes or streak cameras lack the necessary temporal resolution to resolve dynamics at the sub-ps level. Second, many semiconductor systems show stochastic dynamics, where even for nominally identical preparation of the semiconductor system, for repeated experiments its dynamics may vary largely from measurement to measurement. Blinking in quantum dots \cite{Nirmal1996}, spontaneous polariton condensation \cite{Estrecho2018} or cavity feeding \cite{Winger2009} are good examples for semiconductor effects resulting in stochastic dynamics. In these cases, standard optical spectroscopy approaches measure the mean response of the system averaged over all realizations. In order to gain insights into the individual realizations one must perform triggered or conditional experiments. These may either consist of single-shot measurements or photon correlation measurements. The former usually record the response to a single excitation pulse and limit the temporal resolution that may be achieved drastically or provide no temporal resolution at all. The latter correspond to measurements of the conditional dynamics of the system that usually rely again on photo diodes or streak cameras and their limited temporal resolution and accept only one trigger event: the detection of one photon with some well-defined properties chosen beforehand, such as energy or polarization.

One possible way to investigate light fields with enhanced temporal resolution is known from quantum optics, where in contrast to semiconductor physics the properties of a light field itself are of fundamental importance and remarkable progress has been made in demonstrating advanced functionality in the fields of quantum computing \cite{Knill2001}, precision metrology \cite{Ye2008} and secure quantum communication \cite{Gisin2002} under lab conditions. Here, the light fields are used as tools to carry information or to perform precision measurements, so special emphasis is usually placed on a full reconstruction of the states of the light fields \cite{Leonhardt1995} and their spectral-temporal modes \cite{Davis2018}, especially for non-classical light fields such as squeezed states or Schr\"odinger cat states \cite{Raimond1997,Ourjoumtsev2007}. For stationary light fields, a precise characterization of the quantum state may be performed in terms of a tomographic reconstruction via optical homodyne tomography (OHT) \cite{Smithey1993,Lvovsky2009,Breitenbach1997}. However, OHT requires a reference beam with a stable relative phase to the signal of interest. While the duration of this reference beam determines the temporal resolution achievable using OHT and therefore in principle opens up the path towards semiconductor spectroscopy at the sub-ps scale, emission from semiconductors usually consists of photoluminescence, for which no stable phase reference is available. Thus, OHT is usually not applied to semiconductors.   

Here we introduce and demonstrate the technique of non-stationary OHT that makes use of techniques from standard quantum state tomography and adapts them to the special requirements of semiconductor spectroscopy which focuses on non-stationary states, where both amplitude and phase of the signal may vary with time. We realize a time resolution of about 100\,fs, gain access to the phase of the light field without the need for a fixed phase reference and most importantly introduce a postselection mechanism that allows us to measure conditional dynamics that correspond to triggering on events that cannot be triggered upon by conventional means. We demonstrate its capabilities by applying postselective triggering to a light field that shows stochastic dynamics on the sub-picosecond scale: thermal light. In detail, by reconstructing its Wigner function we show that the postselected state consists of a hidden second-order coherent state at any given instant that is masked by inhomogeneous broadening and performs a random walk in phase space on the femtosecond timescale \cite{Siegman1986}. We anticipate our work to open up the way for phase-resolved studies in semiconductor spectroscopy and conditional studies that allow deeper insights into the dynamics of semiconductor systems compared to what has been possible so far.



\section{Theory}
Both stationary and non-stationary OHT are based on balanced homodyne detection (BHD) of light fields \cite{McAlister1997,Roumpos2013}, which requires a local oscillator (LO) that interferes with the signal of interest on a beamsplitter. Then even for weak signal light fields down to the single photon level \cite{Qin2015}, a subsequent balanced photodetector yields an output proportional to the signal field quadrature $q$ in phase with the LO. Figure \ref{fig:setup-husimi-wigner}(a) illustrates such a 4-port OHT assembly in the shaded area titled ``target''. Here, the relative phase $\theta$ between the pulsed LO and the signal light field can be controlled by a phase shifting piezo mirror, as long as $\theta$ is time independent. Then, it is possible to reconstruct the complete signal quantum state in form of its density matrix $\rho$ or its Wigner function from a sufficient number of data points $(q,\theta)$.

Signals that fulfill this criterion, however, are rare. Even laser light is typically not sufficient, since lasers are subject to gain and loss. Thus, their phase slowly randomizes. Stationary OHT experiments routinely overcome this problem by deriving signal and LO from the same laser so that both the signal and the LO show the same phase drift and their relative phase $\theta$ is stable \cite{Wu1986}. For signal light generated independently from the LO, $\theta$ becomes time-dependent and stationary OHT provides only the phase-averaged signal quantum state. In the limit of slow signal phase drift and long coherence times, as is common in quantum optics, the LO may still be actively phase-locked to the signal. In ultrafast semiconductor spectroscopy, where signal coherence times may reach the sub-picosecond regime and phase dynamics may change on the same timescale, active phase-locking is rarely an option. 

\begin{figure}
	\centering
	\includegraphics[width=0.6\columnwidth]{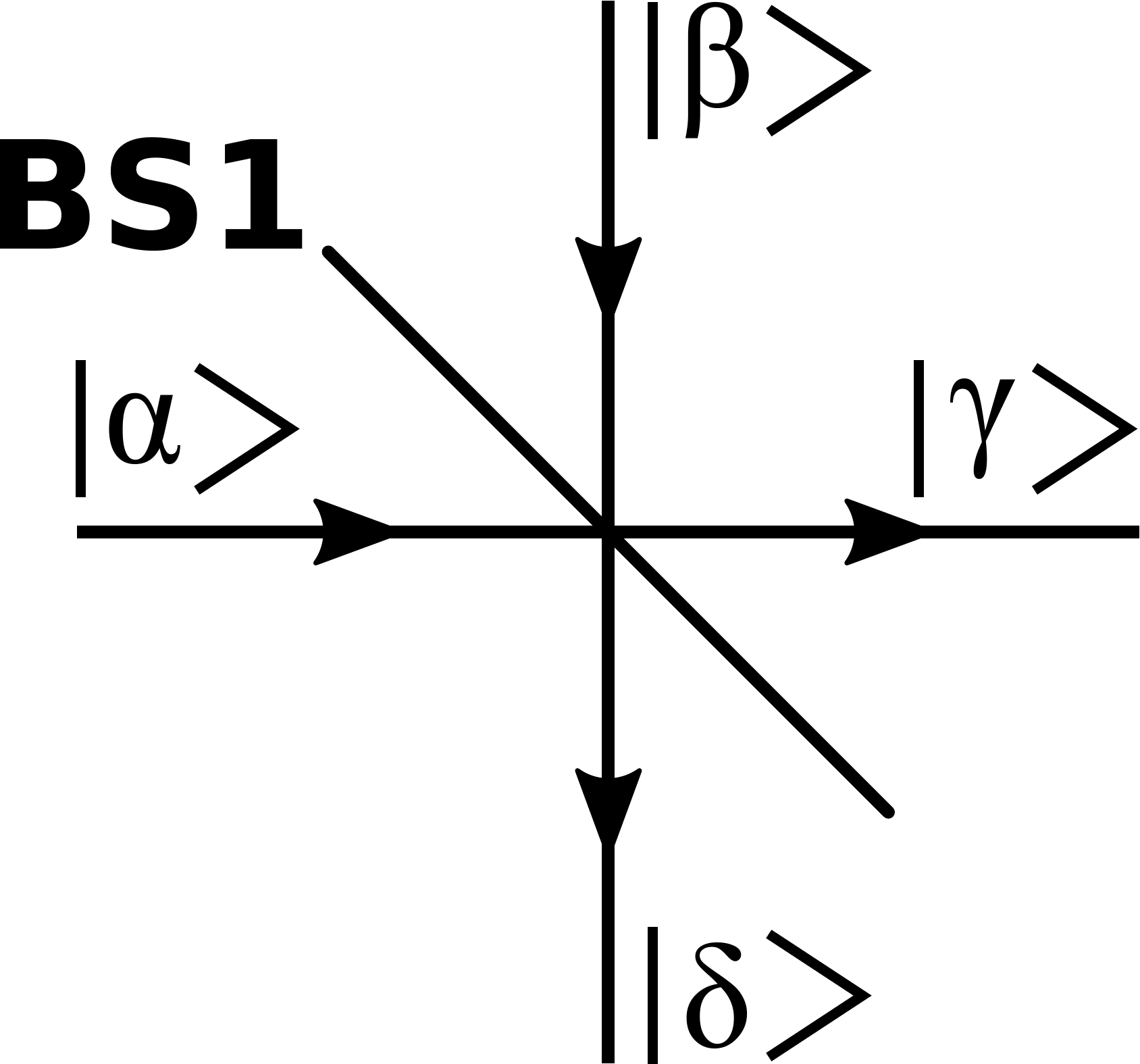}
	\caption{Incoming ($|\alpha\rangle$, $|\beta\rangle$) and outgoing ($|\gamma\rangle$, $|\delta\rangle$) coherent states at a lossless beamsplitter.}
	\label{fig:BS1}
\end{figure}

\begin{figure}
	\centering
		\includegraphics[width=0.9\columnwidth]{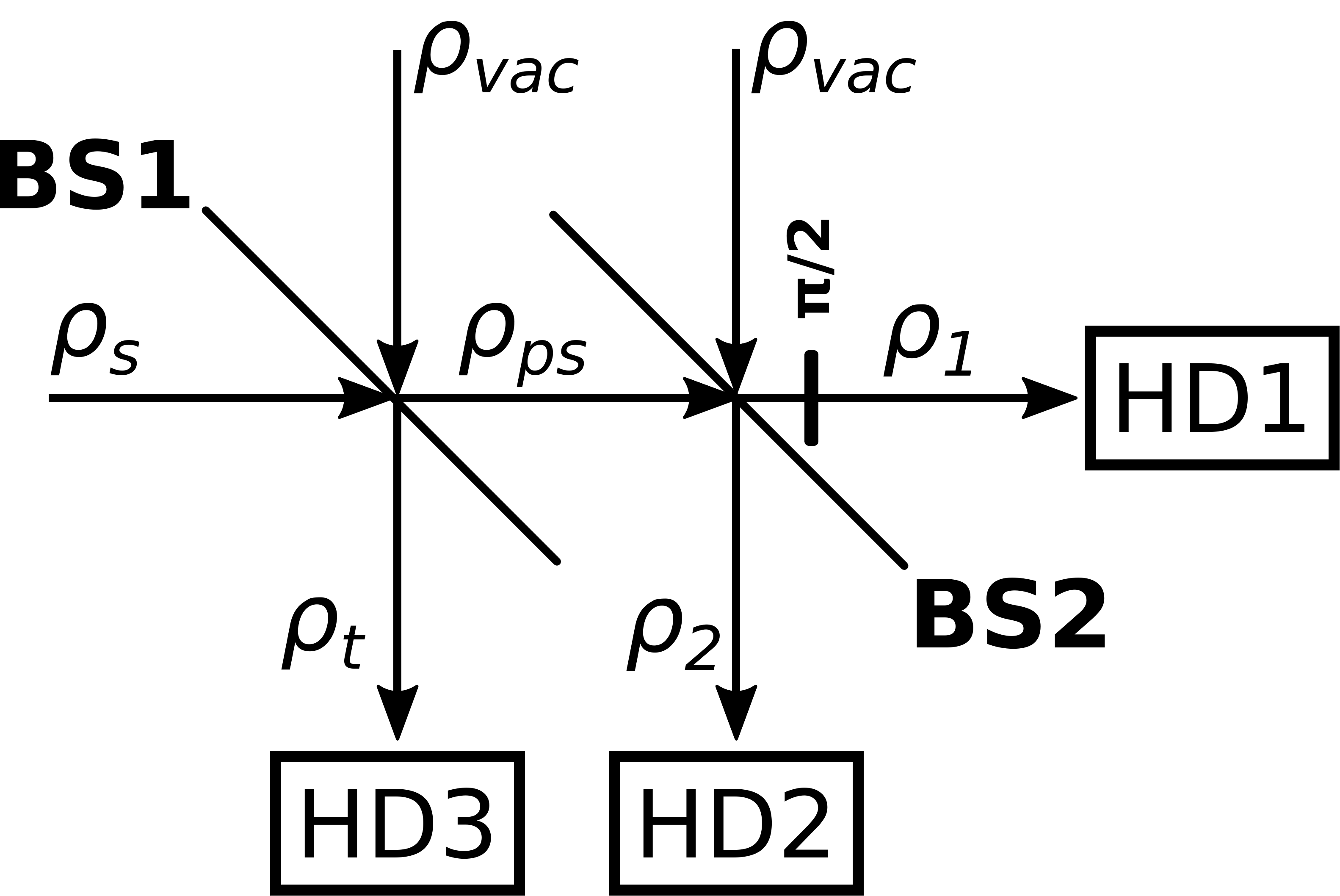}
	\caption{$12$-port homodyne detector for non-stationary optical homodyne tomography. The letter $\rho$ indicates the density operator $\hat{\rho}$ of the corresponding quantum state. The signal quantum state $\hat{\rho}_{s}$ is distributed among the $4$-port homodyne detectors HD1, HD2 and HD3 by the two beamsplitters BS1 and BS2. Further details are discussed in the text.}
	\label{fig:twoBeamsplitters}
\end{figure}

With non-stationary OHT, we aim at capturing the time dependence of the relative phase $\theta$ explicitly. This requires information about the signal light field at some instant $t_0$ and after some delay $\tau$. Therefore, we split the signal into two parts using a beamsplitter (see figure \ref{fig:BS1}); one part is detected in the target arm, which contains a single balanced homodyne detector that samples the signal at time $t_0 + \tau$. The other part of the setup is called the postselection arm and consists of two balanced homodyne detectors that sample a part of the signal field at orthogonal quadratures at time $t_0$ as shown in figure \ref{fig:twoBeamsplitters}. This part is effectively an 8-port OHT setup \cite{Qi2020}, which allows us to measure the Husimi-Q phase space function \cite{Schleich2001} of the light field and to estimate the relative phase $\theta$ between LO and signal at $t_0$. A key point is that the light fields arriving at the target and postselection arms are usually not independent of each other. As they emanate from the two output ports of the same beamsplitter, they are given by the beamsplitter transform applied to the states at the input ports as depicted in figure \ref{fig:BS1}. In the following we assume that the transmission and reflection coefficients of the beamsplitter are given by $t$ and $r$, respectively. Since each density operator $\hat{\rho}$ can be expanded into coherent states $|\alpha\rangle$ with the help of the Glauber-Sudarshan P-function by
\begin{equation} \label{eqn:P-rho}
	\hat{\rho} = \int P(\alpha)\,|\alpha\rangle\langle\alpha|\,d^2\alpha,
\end{equation}
the transformation of any input quantum states on a beamsplitter into output quantum states can be understood by only considering coherent states \cite{Schleich2001}. The combined input quantum state of the beamsplitter for a fixed spatiotemporal mode is then given by:
\begin{equation}
	\hat{\rho}_{in} = \int \int P_{s1}(\alpha) P_{s2}(\beta) |\alpha\rangle\langle\alpha| \otimes |\beta\rangle\langle\beta| \,d^2\beta d^2\alpha,
\end{equation}
where $s1$ and $s2$ denote the quantum states of the two light fields at the input ports of the beam splitter. For simplicity, we may also define a joint two-mode P-function $P(\alpha,\beta)=P_{s1}(\alpha)P_{s2}(\beta)$. The joint quantum state at the beam splitter output ports is then given by
\begin{align}
	\hat{\rho}_{out} = \int \int &P(\alpha,\beta) |t \beta + r \alpha\rangle\langle t \beta + r \alpha| \nonumber \\
	&\otimes |t \alpha - r \beta\rangle\langle t \alpha - r \beta|\,d^2\beta\,d^2\alpha.
\end{align}
$\hat{\rho}_{out}$ can also be represented in terms of the output variables $\gamma = t \alpha - r \beta$ and $\delta = t \beta + r \alpha$:
\begin{equation} \label{eqn:rho-out}
	\hat{\rho}_{out} = \int \int P(t \gamma + r \delta,t \delta - r \gamma) |\delta\rangle\langle\delta| \otimes |\gamma\rangle\langle\gamma|\,d^2\gamma\,d^2\delta.
\end{equation}
Obviously, the beamsplitter mixes the input states such, that the output states are usually not independent of each other but may be correlated to some degree depending on the states of the light fields at the input ports.\\
As the amplitudes of coherent states are complex-valued, the joint two-mode P-function depends on four parameters in total. In order to perform measurements on the joint output state, one may now perform balanced homodyne detection at the different output ports of the beam splitter simultaneously. It is usually not possible to measure the P-function directly. Therefore, real experiments will instead measure other experimentally accessible phase space distributions, such as the Wigner function $W(\alpha,\beta)$ or the Husimi Q function $Q(\alpha,\beta)$, which are connected to the P-function via convolutions \cite{Cahill1969}:
\begin{align}
	Q(\alpha,\beta) = \frac{1}{\pi^2} \int \int &P(\alpha,\beta)\exp\,(-|{\alpha-\alpha^\prime}|^2) \nonumber\\
	& \times\exp\,(-|{\beta-\beta^\prime}|^2)\,d^2\beta^\prime\,d^2\alpha^\prime \\
	W(\alpha,\beta) = \frac{1}{\pi^2} \int \int &P(\alpha,\beta)\exp\,(-2|{\alpha-\alpha^\prime}|^2) \nonumber\\
	&\times\exp\,(-2|{\beta-\beta^\prime}|^2)\,d^2\beta^\prime\,d^2\alpha^\prime.
\end{align}
One possible approach to perform such joint measurements is a 12-port homodyne detector setup placed at the output ports of the beam splitter shown in figure \ref{fig:BS1}, where an 8-port OHT setup forms the postselection arm of the experiment as outlined above, while a 4-port OHT setup forms the target arm. The former allows one to measure Husimi Q-functions, while the latter is a standard setup to measure Wigner functions.

While equation (\ref{eqn:rho-out}) is general and valid for arbitrary input fields, here we are interested mainly in demonstrating that we are able to perform conditional measurements with sub-ps temporal resolution. As thermal light is known to show fluctuations on such short time scales, we will focus on a certain combination of input states given by a thermal signal light field and a vacuum state and discuss the postselection procedure using this example.

 Figure \ref{fig:twoBeamsplitters} shows a simplified sketch of this configuration, where the first two balanced detectors correspond to the postselection arm, while the third balanced detector corresponds to the target arm.

The joint P-function is a product of a thermal signal state and a vacuum P-function:
\begin{equation}
	P(\alpha,\beta) = \frac{1}{\pi \bar{n}} \exp\,(-\frac{|{\alpha}|^2}{\bar{n}}) \delta\,(\beta),
\end{equation}
where $\bar{n}$ denotes the mean photon number of the thermal signal state. This corresponds to the following Q-function:
\begin{equation}
	Q(\alpha,\beta) = \frac{1}{\pi^2 (1+\bar{n})} \exp\,(-\frac{|{\alpha}|^2}{1+\bar{n}}) \exp\,(-|{\beta}|^2) ,
\end{equation}
which may be expressed in the output variables $\delta$ and $\gamma$ as follows:
\begin{align}\label{eqn:jointQ}
Q(\gamma,\delta) = &\frac{1}{\pi^2 (1+\bar{n})} \exp\,(-\frac{|{t \gamma + r \delta}|^2}{1+\bar{n}})\nonumber\\
 &\times\exp\,(-|{t \delta - r \gamma}|^2).
\end{align}

Starting from this equation, we may discuss the continuous variable equivalent of the conditional discrete variable photon correlation measurements mentioned earlier: While such measurements record the conditional probability to detect a photon at time $t_0 + \tau$ triggered on a photon detection event at time $t_0$, we record the conditional Wigner function of the light field overlapping with the LO mode in the target arm at a time $t_0 + \tau$ triggered on the light field overlapping with the LO in the postselection arm having some well defined values of phase and amplitude at time $t_0$. We may choose these amplitudes and phases freely. This provides us with a wider choice of triggering conditions compared to what is possible in photon correlation measurements. The simplest postselection condition consists of picking a single amplitude and phase $\gamma=\gamma_{ps}$ in the postselection arm and filtering on the subset of experimental outcomes that show this value. Inserting this value into equation (\ref{eqn:jointQ}) yields the expected conditional Q-function $Q_c(\delta)$ of the light field in the target arm. After normalization it reads:
\begin{align}
Q_c(\delta) =& \frac{1+\bar{n}t^{2}}{\pi(1+\bar{n})} \exp\left(\frac{\left| \gamma_{ps} \right|^{2}}{1+\bar{n}t^{2}}\right)\exp\left(-\left| t \delta - r\gamma_{ps} \right|^{2}\right)\nonumber\\
 &\times\exp\left(-\frac{\left| t \gamma_{ps} + r \delta \right|^{2}}{1+\bar{n}}\right).
\end{align}
One may measure this function by placing another 8-port OHT setup in the target arm. However, due to the often limited number of detection channels that may be read out simultaneously at high speed, it may be more advantageous to place only a 4-port OHT setup in the target arm, which may be used to measure the conditional Wigner function $W_c(\delta)$ instead, which may be calculated from $Q_c(\delta)$ by a deconvolution. Replacing the complex-valued $\delta$ by the adequately scaled quadratures $q$ and $p$, using the commutator relation convention $[\hat{q},\hat{p}]=i$ and introducing the mean photon numbers $\bar{n}_{ps} = \bar{n}t^2$ in the postselection arm and $\bar{n}_t = \bar{n}r^2$ in the target arm before postselection yields
\begin{align}
W_c &\left(q,p\right) = \frac{1+\bar{n}_{ps}}{\pi\left(1+\bar{n}+\bar{n}_t\right)}\nonumber \\
&\times\exp\left(-\frac{\left(\left(1+\bar{n}_{ps}\right)q-\sqrt{2 \bar{n}_{ps} \bar{n}_{t}}\,q_{ps}\right)^{2}}{\left(1+\bar{n}_{ps}\right)\left(1+\bar{n}+\bar{n}_t\right)}\right)\nonumber \\
 & \times \exp\left(-\frac{\left(\left(1+\bar{n}_{ps}\right)p-\sqrt{2 \bar{n}_{ps} \bar{n}_{t}}\,p_{ps}\right)^{2}}{\left(1+\bar{n}_{ps}\right)\left(1+\bar{n}+\bar{n}_t\right)}\right)
\label{eq:targetWigner}
\end{align}
for the conditional Wigner function of the light field in the target arm.

Depending on the nature of the initial light fields, one may gather different kinds of information from this postselected state. While the conditional Wigner function will also be of interest for entangled states or non-classical photon number states, in the following we focus on the typical case in spectroscopy, where one usually encounters semiclassical Gaussian pure or mixed states, such as coherent or thermal light fields. Here, we would like to point out that the spectroscopic implementation of quantum state tomography is very different from the one used for identification of mode profiles of quantum states in quantum optics. For time-resolved quantum state tomography of e.g. Fock states, any mismatch between the spatial, spectral, temporal or polarization modes of the signal and the local oscillator will lead to enhanced vacuum contributions \cite{Ogawa2016,Morin2013,Qin2015,Polycarpou2012}. Accordingly, usually one aims at having perfect overlap between the LO and the signal in all properties mentioned above. 
In spectroscopy, one is instead interested in filtering the signal to single out the individual spectral, spatial, temporal and polarization modes. As only the part of the signal that overlaps with the LO is detected, we may achieve exactly that by tailoring the properties of the LO accordingly. For mixed states, this corresponds to postselective filtering on a subset of states with weights given by their overlap with the LO mode. The temporal resolution we may achieve for quadrature measurements is limited only by the duration of the LO pulse, which amounts to about 100\,fs in our case.

Here, we choose to investigate the same mode in the target and postselection arms of our setup. For Gaussian states, such as the case of the thermal light field outlined above, there is a well defined phase and amplitude relationship between the fields in the two arms. Accordingly, postselection on a single quadrature value will result in a certain amplitude, phase and variance of the target light field for the same mode. Accordingly, the amplitudes and variances of postselected states may serve as a benchmark for how well our implementation of the continuous variable postselection procedure works.

To compute the postselected variance $Var_c(q)$ of the conditional target quantum state along one quadrature axis $q$, it is necessary to calculate the expectation values $\langle q \rangle_c$ and $\left< q^2 \right>_c$. These read:
\begin{equation}
	\left< q \right>_c = \int_{-\infty}^{\infty}\int_{-\infty}^{\infty}W_c(q,p)\,q\,dp\,dq = \sqrt{2}q_{ps}\frac{\sqrt{\bar{n}_{ps}\bar{n}_{t}}}{1+\bar{n}_{ps}}
	\label{eq:q_t}
\end{equation}
\begin{align}
	\left< q^{2} \right>_c =& \int_{-\infty}^{\infty}\int_{-\infty}^{\infty}W_c(q,p)\,q^{2}\,dp\,dq \nonumber\\
	=& \frac{1+\bar{n}+\bar{n}_{t}}{2\left(1+\bar{n}_{ps}\right)}+\langle q\rangle_c^2. \label{eq:q2_t}
\end{align}
In the following, we will introduce our experimental implementation of the postselective 12-port homodyne detection setup and confirm its functionality by verifying that the properties of the postselected state match the predictions given above for a thermal input field. Afterwards, we will deliberately delay the LO in the target arm to sample a different temporal mode compared to the postselection condition. Doing so allows us to track the dephasing of the postselected state.

\section{Experimental Setup}
\begin{figure*}[hbt]
	\includegraphics[width=0.9\textwidth]{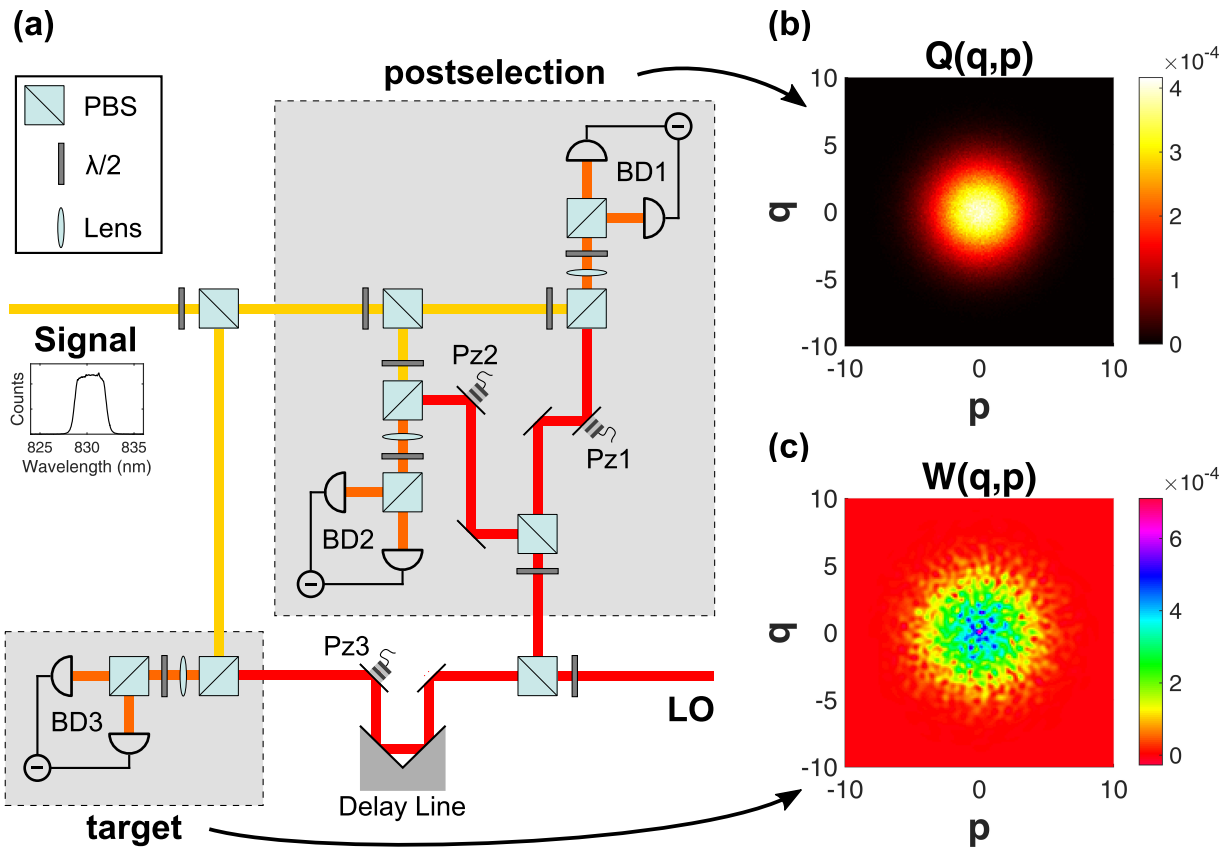}
	\caption{In our 12-port homodyne detection setup (a), a strong local oscillator (LO) and a signal beam are each split into three beams and are interfered on polarizing beamsplitters (PBS) before being detected by three balanced detectors (BD). While BD1 and BD2 are placed in the postselection arm in order to measure the Husimi-Q function (b) of the thermal signal light field, BD3 is placed in the target arm which is used to reconstruct its Wigner function (c). On average, there were $\bar{n}_t=5.63$ photons in the target arm and $\bar{n}_{ps}=11.86$ photons in the postselection arm. The inset provides the signal's spectral density. Pz: piezo mirror. $\lambda/2$: half-wave plate.\label{fig:setup-husimi-wigner}}
\end{figure*}

Figure \ref{fig:setup-husimi-wigner}a illustrates the experimental realization of our non-stationary OHT setup. The target arm was realized by a 4-port OHT assembly that features a tunable delay relative to the 8-port homodyne detector forming the postselection arm. The thermal signal light originated from a Toptica DL pro continuous wave diode laser operated far below threshold. It was filtered spectrally using an optical bandpass with a specified center wavelength of $830\,$nm and a full width at half maximum of $3\pm1\,$nm. The resulting spectral density of the signal light is shown in Fig.~\ref{fig:setup-husimi-wigner}a. The pulsed LO was generated by a Ti:sapphire laser operated at a repetition rate of $75.4\,$MHz with a pulse duration of about $120\,$fs. Its central wavelength was set to $830\,$nm and the full width at half maximum of its Gaussian shaped spectral density was about $10\,$nm. The homodyne detectors used had a bandwidth of $100\,$MHz, a transimpedance gain of $5000\,$V/A, and were provided by Femto (Model HCA-S). The voltage signal from each detector was processed by band-stop filters at $75.4\,$MHz and $150.8\,$MHz and a $100\,$MHz low-pass filter to remove all harmonics of the laser repetition rate. After that, it was amplified by a factor of $5$ with a $300\,$MHz SR455 voltage amplifier from Stanford Research Systems before being digitized by a $5\,$GS/s analog-to-digital converter M4i.2234-x8 from Spectrum providing a sampling rate of $1.25\,$GS/s per detection channel \cite{Lueders2018}. To calculate the quadrature amplitudes $q_1$, $q_2$ and $q_3$ for each homodyne detector, the acquired voltage signals were first integrated for each laser pulse separately. After removing spurious correlations between consecutive quadrature measurements \cite{Kumar2012}, the result was normalized with respect to the LO to meet $Var(q_i)=1/2$ for the vacuum state. This corresponds to the commutator convention $[\hat{q},\hat{p}]=i$ \cite{Lvovsky2009} for $\hbar=1$. The detectors used standard Hamamatsu S5972 photo diodes with a peak quantum efficiency of about 84\,$\%$. We aim mostly at classical light fields, where losses due to limited efficiency are not crucial. Still, if a precise treatment of losses and finite quantum efficiency becomes mandatory for some system under study, it can be incorporated into the maximum likelihood algorithm using standard procedures \cite{Lvovsky2004}.

For postselection, we needed to identify the relative phase $\phi_{ps} = \arctan(q_{ps}/p_{ps})$ between LO and signal in the postselection arm. To this end, the piezo actuators Pz1, Pz2 and Pz3 were driven with frequencies of $50\,$Hz, $0.5\,$Hz and $0\,$Hz, respectively. The voltages were modulated sinusoidally with a maximum travel range of $2\,\mu$m. The first postselection step ensured that the postselected quadratures $q_1$ and $q_2$ represent orthogonal measurements by calculating the smoothed time dependence of $q_1 \cdot q_2$ and selecting on a narrow range around zero amplitude. Additionally, the smoothed time dependence of $q_1\cdot q_3$ was used to calculate the relative phase $\phi_t$ between the LOs of the first and third homodyne detectors. The resulting dataset $(q_t,\phi_t,q_{ps},p_{ps})$ consisted of the postselected quadrature and phase values $(q_t,\phi_t)$ in the target arm and the orthogonally measured quadratures $(q_{ps},p_{ps})$ in the postselection arm. The relative phase $\theta$ of the LO and the signal light in the target arm could then be retrieved by combining the relative phase $\phi_{ps}=\arctan(q_{ps},p_{ps})$ between LO and signal in the postselection arm with $\phi_t$. This relative phase was calculated with the four-quadrant inverse tangent function, resulting in a dataset consisting of $(q_t,\theta,q_{ps},p_{ps})$ for each individual LO pulse. The dataset used for the results shown in Fig.~\ref{fig:postselection}, for instance, consisted of $2\,309\,377$ quadruples of this form. By measuring the signal's Q-function (Fig.~\ref{fig:setup-husimi-wigner}b), we also gained access to its instantaneous amplitude $A_{ps}=\sqrt{q^2_{ps}+p^2_{ps}}$. Thus, postselecting on specific parts of the Q-function gave us the ability to postselect a target quantum state at $t_0$ and to monitor the conditional dynamics of this postselected state based on our choice of amplitude and phase. To this end, postselecting means that we investigated all of the measured quadruples $(q_{ps},p_{ps},q_t,\theta)$, filtered for a certain range of $q_{ps}$ and $p_{ps}$ and only kept the corresponding $q_t$ and $\theta$. We would like to emphasize that this was done fully during the data analysis stage and the postselected values do not need to be chosen before the measurement.

\section{Conditional Quantum State Tomography}
To demonstrate the capabilities of the conditional OHT technique, we investigate thermal light as the signal state because both its amplitude and phase exhibit ultrafast stochastic dynamics on the femtosecond timescale and the possible time resolution we can achieve is the most important figure of merit for semiconductor spectroscopy. Therefore, thermal light is a good benchmark for the capability of our technique to single out the instantaneous state of a light field from the full set of possible amplitudes and phases the non-stationary state may take.

We performed two types of measurements. First, we validated the postselection protocol and identified its fundamental noise limit at zero delay. Second, we investigated the temporal evolution of conditional target quantum states for several choices of $q_{ps}$ and $p_{ps}$ as will be shown in the next section.

First, for comparison standard unconditional Wigner functions $W(q,p)$ without performing any postselection are reconstructed from target quadratures $(q_t,\theta)$ with the help of a maximum likelihood algorithm \cite{Lvovsky2004,Lvovsky2009}. Performing standard stationary OHT in the target arm, we find the typical Wigner function of a thermal state (Fig.~\ref{fig:setup-husimi-wigner}c).

\begin{figure*}[hbt]
	\includegraphics[width=0.9\textwidth]{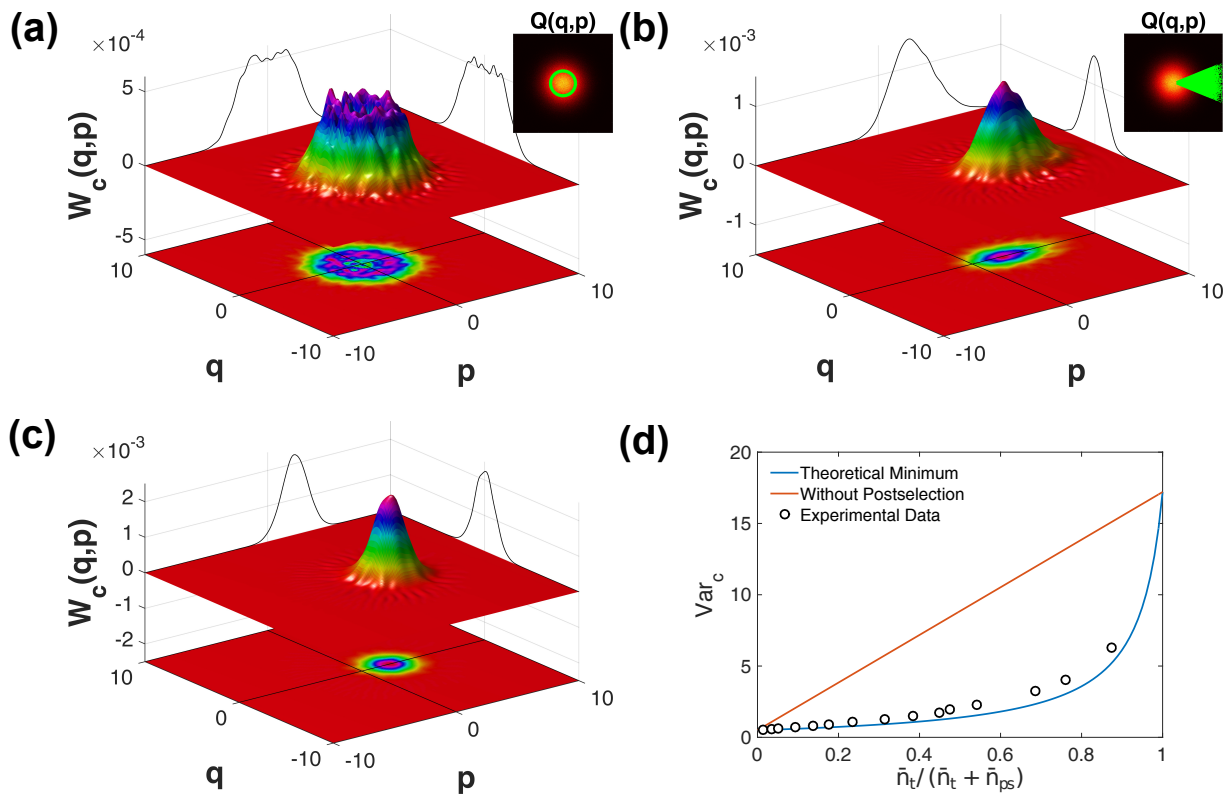}
	\caption{Conditional Wigner functions when postselecting on (a) amplitude and (b) phase of the Husimi-Q function (selected regions are illustrated in the inset). When selecting on the amplitude and using the available phase information at the same time for the reconstruction (c), different photon numbers $\bar{n}_t$ and $\bar{n}_{ps}$ in the target and postselection arms result in different average variances (d) of the conditional Wigner function. The graphs on the walls of (a), (b) and (c) are integral projections of the shown Wigner functions scaled to have the same maximum values as the Wigner functions. The colors correspond to elevation. \label{fig:postselection}}
\end{figure*}

Next, we tested whether our postselection protocol works as expected and allows us to postselect on arbitrary combinations of amplitude and phase of the light field. To this end, we chose a time delay of $\tau=0$, so that both the target and the postselection arm measure the signal light field at the same time. In order to obtain conditional Wigner functions $W_c(q,p)$, the same Wigner function reconstruction procedure as before was applied, but out of all measured quadrature quadruples $(q_{ps},p_{ps},q_t,\theta)$ only those quadruples containing a selected range of $(q_{ps},p_{ps})$ in the postselection arm were chosen. The reconstruction of the Wigner function was performed only on the corresponding subset of $(q_t,\theta)$. As a first example, we chose a ring-like subset of amplitudes such that $A_{ps}=\sqrt{q_{ps}^2+p_{ps}^2}=2.5 \pm 0.25$ without any phase postselection. The expected state of the conditional light field in the target arm is then also a ring-like quadrature distribution with a slightly smaller conditional amplitude of $A_c=2.25$. The expected value is determined by $A_{ps}$ and the unselected mean photon numbers $\bar{n}_t$ and $\bar{n}_{ps}$ in the target and postselection arms as given by equation (\ref{eq:q_t}):
\begin{equation}
	A_c = \sqrt{2} A_{ps} \frac{\sqrt{\bar{n}_{ps}\bar{n}_t}}{1+\bar{n}_{ps}}.
	\label{eqn:AtFromAps}
\end{equation}
The actually measured conditional target Wigner function is a narrow circular distribution with an average amplitude of $\left<A_c\right>=2.247$ as shown in Fig.~\ref{fig:postselection}a. This outcome shows excellent agreement with the expected result and demonstrates that we are indeed able to perform postselection on the amplitude of the light field. Along the same lines, Fig.~\ref{fig:postselection}b shows the conditional target Wigner function $W_c$ when postselecting on a range of phases $\phi_{ps} = \arctan(q_{ps}/p_{ps}) = 0 \pm \pi/8$, which is a narrow distribution around the positive part of the $p$-axis. $W_c$ also shows a distribution that occupies a narrow region of about $\pm \pi/8$ around the positive $p$-axis, which demonstrates that we may also perform postselection on the phase of the light field. Thus, we gained control over amplitude and phase of the conditional target quantum state by postselecting on the Q-function.

When postselecting on a narrow range of both amplitude and phase of a thermal state, the target arm Wigner function resembles a coherent state (see Fig.~\ref{fig:postselection}c). However, the quadrature distribution of the conditional Wigner function is slightly broader than the standard Wigner function of an ideal coherent state since our postselection protocol requires to split the signal quantum state in two parts, which adds vacuum noise depending on the splitting ratio. The theoretical limits for the conditional target quadrature variances $Var_c(q_t)$ and $Var_c(p_t)$ that may be achieved for the conditional Wigner function after postselection on a single amplitude and phase of a thermal state can be expressed in terms of $\bar{n}_t$ and $\bar{n}_{ps}$ as given by equations (\ref{eq:q_t}) and (\ref{eq:q2_t}):
\begin{equation} \label{eqn:limit}
	Var_c(q_t) = Var_c(p_t) = \frac{1+\bar{n}+\bar{n}_t}{2(1+\bar{n}_{ps})}
\end{equation}
The lower bound of Eq.~(\ref{eqn:limit}) is the variance $1/2$ of a vacuum state for $\bar{n}_t\ll\bar{n}_{ps}$ while its upper bound is given by the variance $\bar{n}+1/2$ of the original signal quantum state for $\bar{n}_t\gg\bar{n}_{ps}$. Therefore, one may interpret the postselected conditional state of the thermal light field as a coherent state mixed with additional vacuum noise determined by the signal splitting ratio. Figure \ref{fig:postselection}d compares the experimentally determined quadrature variances for different splitting ratios with the limit given by Eq.~(\ref{eqn:limit}). They are in good agreement. Some small residual deviations may result from imperfections in the experimental setup. Please note that similarly increased quadrature variances would also mask the non-classicality of non-classical states, which are, however, only rarely relevant for semiconductor spectroscopy.

\begin{figure}[hbt]
	\centering
	\includegraphics[width=0.95\columnwidth]{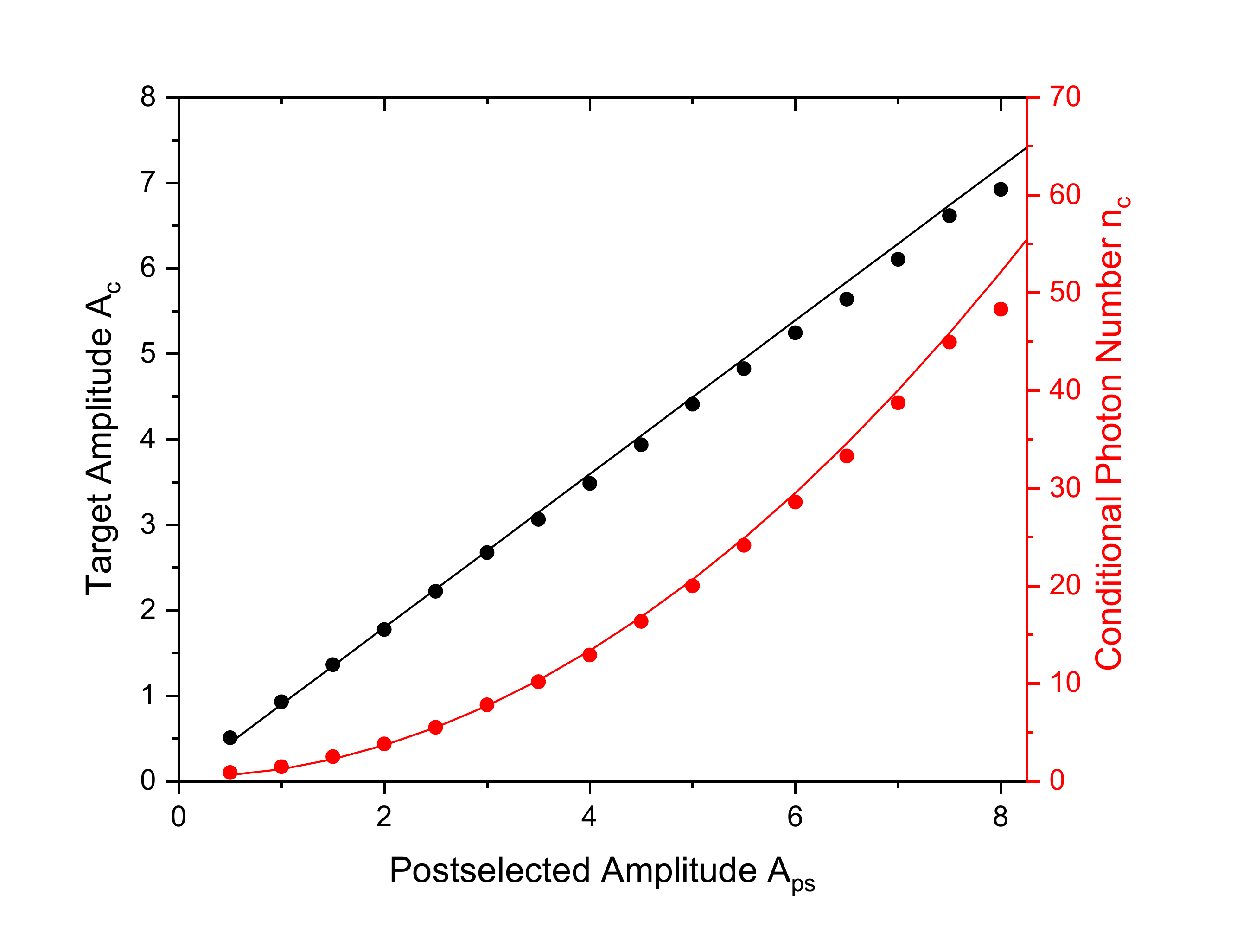}
	\caption{Verification of the postselection protocol. Measured conditional target amplitudes $A_c$ (black dots) and conditional photon numbers $n_c$ (red dots) for several postselected amplitudes $A_{ps}$ at zero delay. The solid lines give the values expected from theory.}
	\label{fig:ps-proof}
\end{figure}

As a further demonstration that the postselection works as intended, we took the same set of data as before and evaluated a series of conditional target arm Wigner functions for postselecting on linearly increasing quadrature values. The variance of the postselected state should be given by equation (\ref{eqn:limit}) and its photon number is expected to be
\begin{equation}
	n_c = A^2_c + Var_c(q_t)-\frac{1}{2}.
	\label{eq:amplitudes}
\end{equation}
Figure \ref{fig:ps-proof} shows the results of these measurements. Both the conditional amplitude and the conditional photon number are in good agreement with the theoretical expectations. This results also shows that our experimental setup works as intended.

\section{Non-stationary Conditional Quantum State Tomography}
The measurements discussed so far show that we are able to postselect and effectively trigger on the instantaneous quadratures of the light field in the postselection arm. The validity of the postselection scheme has also already been demonstrated in other contexts, e.g. by demonstrating an eavesdropping attack on a trusted quantum random number generator \cite{Thewes2019}. We will now demonstrate the central advantage of our experimental technique: The possibility to perform conditional spectroscopy which allows us to perform postselection on arbitrary amplitudes and phases at some instant $t_0$ and observe the conditional dynamics of the postselected selected state after and even before this time. 

\begin{figure*}[hbt]
	\includegraphics[width=0.8\textwidth]{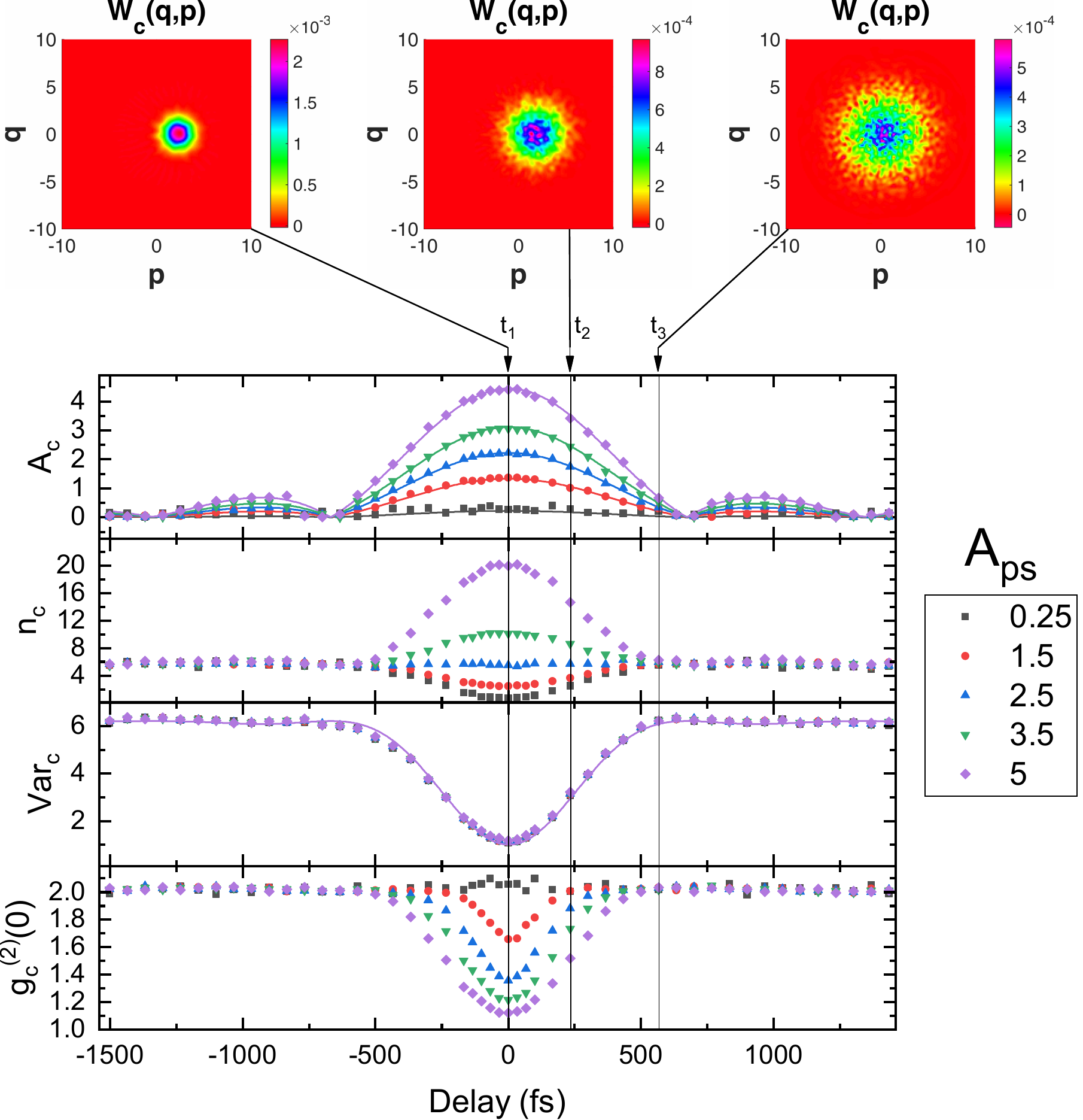}
	\caption{Time dependence of the conditional target amplitude $A_c$, photon number $n_c$, variance Var$_c$, and equal-time second order correlation function $g_c^{(2)}(0)$ of the conditional quantum state together with snapshots of three representative Wigner functions. The solid lines are the theoretically expected dependencies derived from the spectral density function in Fig.~\ref{fig:setup-husimi-wigner}(a), following the Wiener-Khintchine theorem \cite{Mandel1995}. \label{fig:timeseries}}
\end{figure*}

To this end, we track the evolution of the light field in the target arm in time by sampling it at different time delays $\tau$ with respect to the postselection arm, while maintaining the same postselection criteria. We will show that the arising dynamics may be interpreted as dephasing due to inhomogeneous broadening. Figure \ref{fig:timeseries} shows conditional Wigner functions of the light field in the target arm at three characteristic times. It also presents the time evolution of four key parameters for postselecting on several different amplitude values $A_{ps}$ of the light field in the postselection arm at time $t_0$: the conditional amplitude, photon number, quadrature variance and equal-time second-order photon correlation function. These results demonstrate unambiguously that we are able to single out individual realizations in a system undergoing stochastic dynamics and resolve the corresponding conditional dynamics.

We now analyze the information we gain about the stochastic dynamics of the thermal light field by performing conditional spectroscopy shown in figure \ref{fig:timeseries} in detail. The first quantity of interest is the decay of the amplitude of the conditional light field. The mean amplitude of any thermal light fields amounts to zero and therefore the decay of the conditional signal amplitude towards this value carries information about the stochastic dynamics of the system. Indeed, the conditional target amplitude $A_c(\tau)$ follows the Fourier transform of the signal spectral density scaled for $A_c(0)$ and shows a Gaussian decay. At large delays of around 1\,ps a small revival of the target amplitude can be observed due to a part of the modes coming into phase again. Both observations are consistent with interpreting the thermal state as consisting of a set of modes, where the overall coherence time is determined by the dephasing time of the inhomogeneously broadened ensemble, but each subensemble shows a longer phase coherence time. Accordingly, at each instant the coherent sum of all the subensembles will resemble a coherent state moving in phase space. This assumption is backed by the conditional variance $\left<Var_c(q_t)\right>$ of the target quadrature distribution which is independent of $A_{ps}$, as predicted by Eq.~(\ref{eqn:limit}). Its value at zero delay always matches the expected width of a coherent state subject to additional vacuum noise. 

Next, we show that we may also trace the conditional photon number of the light field. Both the mean quadrature amplitude and variance contribute to the mean photon number of a light field. For a coherent state, the first component dominates. This behavior is shared by the instantaneous photon number $n_c(0)$ of the target state: it falls below the steady state of $\bar{n}_t$ for small values of $A_{ps}$ and rises above it for large values of $A_{ps}$. For long delays, in all cases the photon number of the postselected state returns to the mean photon number, which shows that we are able to track conditional photon number dynamics. The steady state for larger delays is then again dominated by the variance due to dephasing in phase space. 

Finally, one of the hallmark indicators of the non-stationary nature of a thermal light field is the equal-time second-order correlation function $g^{(2)}(0)$ that measures the relative photon number noise of a light field. As thermal light fields are subject to stochastic dynamics, they show enhanced photon number fluctuations corresponding to $g^{(2)}(0)=2$. Accordingly, any reduction of the equal-time second-order correlation function of the postselected state $g_c^{(2)}(0)$ is a direct indicator of how well we are able to single out a single instantaneous state of the light field from all of the states the system may take. The data clearly shows that large delays or low postselection amplitudes both result in thermal state characteristics with $g_c^{(2)}(0)=2$. This is not surprising. For delays longer than the coherence time of the light field, the state of the light field and its photon number are random again and the light field is again in the non-stationary thermal state, while postselection on small amplitudes essentially selects states with tiny photon number close to the vacuum, where coherent states and thermal states become very similar. 

However, for large $A_{ps}$ and small delays, the conditional state shows significantly reduced photon number noise with $g_c^{(2)}(0)$ down to values of 1.12 which is close to the value of $1$ expected for a coherent state. The remaining difference can be attributed to the additional noise described by Eq.~(\ref{eqn:limit}). Since this contribution is fixed for a given splitting ratio, its influence on $g_c^{(2)}(0)$ decreases for increasing photon numbers. This result shows that it is possible to postselect on well-defined conditional states and supports interpreting this conditional postselected light field as a coherent state.

For a full characterization of the postselected light field state, we also reconstructed its Wigner function. The upper part of Fig.~\ref{fig:timeseries} shows the conditional Wigner function at three different time delays. At all times, it has Gaussian shape and the results directly allow us to monitor the dephasing of the light field (see supplementary video for the complete Wigner function time dependence). The full information contained in this video provides a new postselective continuous variable phase space approach towards studying both the amplitude and phase of signal light fields in semiconductor spectroscopy, which is especially promising for studying stochastic dynamics.

Compared to other experimental techniques, the strength of our approach lies in the fact that the conditional dynamics of the target state may be observed based on any arbitrary combination of amplitude and phase as the postselection condition, even for steady states that show large fluctuations. At current, there are only few experimental techniques that offer similar insights compared to what we can achieve and they are much more limited. The simplest example is pump-probe spectroscopy, which offers a similar temporal resolution \cite{Grosse2014}. However, for pump-probe spectroscopy one does not postselect the state of the system, but rather tries to create it using pump beams of different intensity and monitors the relaxation of the signal. Pump-probe spectroscopy therefore obviously does not yield access to spontaneous processes in steady-state systems and averages over different relaxation pathways. A spectroscopic technique that also yields results for steady state systems is spin-noise spectroscopy \cite{Crooker2004,Oestreich2005}, which works only for spin and operates in the frequency domain. Therefore it yields information about the frequency distribution of noise, but does not allow for direct observation of conditional dynamics. The experimental techniques that are probably closest to our approach are photon correlation measurements, where the conditional dynamics of the system may be observed at different times $t_0+\tau$ conditioned on the detection of a photon at $t_0$. However, the temporal resolution that may be achieved this way is limited to the time resolution of avalanche photo diodes \cite{Ulrich2007, Moody2018} or streak cameras \cite{Assmann2009, Assmann2010} and the postselection condition is limited to the detection of a photon with some energy and polarization of choice, while our approach offers a significantly better temporal resolution and a much wider choice of postselection conditions. We would like to emphasize that just like in photon correlation studies, for non-stationary homodyne tomography the postselection and the target arm do not necessarily need to measure the same mode. As only the part of the signal that is mode matched with the local oscillator is actually measured, preparing the LOs in the two arms with different polarizations as has been done already e.g. in phase-averaged homodyne detection \cite{Blansett2001}, filtering different spectral regions from a spectrally broad LO or using LOs in different orbital angular momentum states opens up the possibility to observe the conditional cross-correlation between these modes. 

\section{Conclusion and Outlook}
In conclusion, we introduced non-stationary homodyne tomography that is a powerful spectroscopic tool to observe conditional dynamics, e.g. in semiconductor systems. We demonstrated its usefulness by monitoring the conditional dynamics of a thermal light field, which is an ideal test system as its ultrafast dynamics are a major challenge in spectroscopy: The conditional dynamics to be observed take place on a time scale of hundreds of femtoseconds and the decoherence process we observed using conditional dynamics is a standard problem in semiconductor physics, where e.g. spin dephasing and other dephasing processes follow a similar dependence.

We want to emphasize that the experimental approach we present here can loosely be considered as a continuous-variable equivalent to a Hanbury Brown-Twiss photon correlation experiment. We measure a four-dimensional joint two-mode function that consists of a joint measurement of the Q function in one arm and the Wigner function in the other arm. We then go on to determine the conditional Wigner function via postselection, which corresponds to a projection of the four-dimensional two-mode function to a two-dimensional single-mode function. Roughly speaking, in our experiment this conditional Wigner function is related to the initial unconditional Wigner function in the same way as the measured value of $g^{(2)}$ in a Hanbury Brown-Twiss experiment is related to the mean photon number of the incoming beam. A major difference between both types of experiments is the range of postselection conditions that may be applied. For a Hanbury Brown-Twiss experiment, the postselection condition is given by detection of a photon with certain properties such as wavelength or polarization. In our experiment, instead a certain range of amplitudes and phases found when performing homodyne detection serves as the postselection criterion. As the range of possible values is continuous in this case, homodyning yields a wider choice of postselection conditions compared to discrete variable photon correlation studies, which may be both an advantage or a disadvantage depending on the light field of interest. Roughly speaking, states that are most easily treated in a Fock state basis lend themselves to photon correlation studies, while states that one will usually describe in a coherent state basis lend themselves to continuous variable experiments. A final notable difference between experiments on conditional dynamics using discrete variable and continuous variable approaches is the role of the vacuum state. As any signal in photon correlation experiments requires the detection of at least one photon as the postselection condition, any vacuum state contributions of the light field of interest are intrinsically filtered in Hanbury Brown-Twiss experiments, while in conditional continuous variable experiments all vacuum state contributions remain present.

As an outlook, we would like to provide some concrete examples for systems where our spectroscopic approach will be beneficial. First, systems that show dynamical bistability or multistability \cite{Fink2018} or fluorescence intermittency \cite{Efros2016} are natural candidates for conditional studies based on continuous variables. At current, these are studied routinely via photon correlation experiments, where the presence of a brighter and a darker state manifests as an enhanced probability to detect further photons if another photon has been detected before. Here, one postselects on photon emission events, but there is no possibility to actively postselect on whether the system is in the bright or in the dark state at some instant. It is just more likely that the photon was emitted, while the system was in the bright state. Being able to postselect on the amplitude of the emitted light field instead directly yields the possibility to postselect on whether the system is in the bright or the dark state or possibly even some grey intermediate state for multistable systems. This higher selectivity will yield more information about multimode systems than photon correlation experiments. Second, systems containing several discernible states could be investigated using our approach. Perhaps the most obvious candidate are lasers showing mode competition, e.g. elliptical micropillar lasers that give rise to two modes of orthogonal polarization coupled to the same gain medium \cite{Leymann2013,Leymann2017}. Here, one may set the LO in the postselection arm to the polarization of one of the modes and the LO in the target arm to the orthogonal polarization. Observing the conditional dynamics of the emission in the target arm, while postselecting on the amplitude in the postselection arm yields direct access to the mode competition dynamics. A third example of systems, where our technique allows for much deeper insights into the physics involved are systems that form coherence spontaneously. A good example is a polariton condensate in a double potential well that acts as a Josephson junction \cite{Lagoudakis2010}. Here, the equivalent of the ac-Josephson effect may occur, which results in the relative phase between the potential wells changing linearly with time. However, the initial phase is chosen randomly, so that standard averaged measurements do not reveal this phase gradient. Using our approach, one may observe the emission from the left potential well in the postselection arm and the emission from the right potential well in the target arm and recover the full phase dynamics by postselecting on certain phases of the light field. Finally, we would like to emphasize that another major strength of our technique is that we record a full set of raw quadratures for every single pulse, so that the whole data analysis can be performed a posteriori in postselection. Due to the laser repetition rate of approximately 75\,MHz, we only require one second to record 75 million sets of three quadratures each, which is fully sufficient to perform data analysis for a single $\tau$. Accordingly, a typical measurement using our approach requires much less measurement time compared to photon correlation measurements using photo diodes or streak cameras.



\begin{acknowledgments}
We thank S.T. Cundiff and M. Kira for stimulating discussions about phase space distributions and A. Lvovsky for discussions about high bandwidth homodyne detector design.\\We gratefully acknowledge support of this project from the DFG via grant number AS 459/1-2 and from the Mercator foundation.
\end{acknowledgments}

\end{document}